\documentclass{article}
\usepackage{spconf,amsmath,amssymb,graphicx}
\usepackage{mathtools}
\usepackage{subfig}
\usepackage{siunitx}
\usepackage{bmpsize}
\usepackage[abs]{overpic}
\usepackage{pgfplots}
\usepackage{adjustbox}
\usepackage[labelsep=period]{caption}
\renewcommand{\thetable}{\Roman{table}}

\usepackage{booktabs,array}
\usepackage{multirow}
\usepackage{balance}
\usepackage{url}

\definecolor{red2}{rgb}{1,0.141176471,0.18823529411}
\definecolor{blue2}{rgb}{0,0.79215686274,1}

\newcommand{\myarrowred}[1][0.1pt]
{   \begin{tikzpicture}[overlay]
	\draw [->,>=stealth,line width=0.5mm,red2] (0, -0.05) -- (0, 0.3);
	\end{tikzpicture}
}

\newcommand{\myarrowblue}[1][0.1pt]
{   \begin{tikzpicture}[overlay]
	\draw [->,>=stealth,line width=0.5mm,blue2] (0, -0.05) -- (0, 0.3);
	\end{tikzpicture}
}
\makeatletter
\renewcommand\footnotesize{%
	\@setfontsize\footnotesize\@ixpt{11.4}%
	\abovedisplayskip 8\p@ \@plus2\p@ \@minus4\p@
	\abovedisplayshortskip \z@ \@plus\p@
	\belowdisplayshortskip 4\p@ \@plus2\p@ \@minus2\p@
	\def\@listi{\leftmargin\leftmargini
		\topsep 4\p@ \@plus2\p@ \@minus2\p@
		\parsep 2\p@ \@plus\p@ \@minus\p@
		\itemsep \parsep}%
	\belowdisplayskip \abovedisplayskip
}
\makeatother

\title{\lowercase{ip}A-M\lowercase{ed}GAN: Inpainting of arbitrary regions in Medical Imaging}
\name{Karim Armanious\textsuperscript{ 1,2}, Vijeth Kumar\textsuperscript{ 1}, Sherif Abdulatif\textsuperscript{ 1}, Tobias Hepp\textsuperscript{ 2}, Sergios Gatidis\textsuperscript{ 2}, Bin Yang\textsuperscript{ 1}}
\address{\textsuperscript{1}University~of~Stuttgart,~Institute~of~Signal~Processing~and~System~Theory,~Stuttgart,~Germany\\
	\textsuperscript{2}University~of~T\"ubingen,~Department~of~Radiology,~T\"ubingen,~Germany
}
\begin{document}
%
\maketitle
\begin{abstract}


Local deformations in medical modalities are common phenomena due to a multitude of factors such as metallic implants or limited field of views in magnetic resonance imaging (MRI). Completion of the missing or distorted regions is of special interest for automatic image analysis frameworks to enhance post-processing tasks such as segmentation or classification. In this work, we propose a new generative framework for medical image inpainting, titled ipA-MedGAN. It bypasses the limitations of previous frameworks by enabling inpainting of arbitrary shaped regions without a prior localization of the regions of interest. Thorough qualitative and quantitative comparisons with other inpainting and translational approaches have illustrated the superior performance of the proposed framework for the task of brain MR inpainting.  
	
\end{abstract}
\vspace{-1mm}
\section{Introduction}
\label{sec:intro}
\vspace{-1mm}
Medical imaging techniques, such as MRI and CT, are an integral part of everyday medical practices. They provide a detailed metabolic and anatomical depiction of organs enabling diagnostics and medication planning. However, causes for deformations in the resultant scans are manifold. For example, motion artifacts result in global deformations in MRI \cite{1}. Additionally, local distortions are common phenomena due to multiple reasons such as selective reconstruction of data, limited fields of view or the superposition of foreign bodies. 

Of course, the information contained within the locally distorted regions is lost in a diagnostic sense. Nevertheless, auto-completion of medical images via inpainting would enable the utilization of corrupted scans in post-processing tasks \cite{2}. For this purpose, only the global image properties are of interest instead of the detailed diagnostic information. For example, completion of localized distortions in MR scans due to metallic implants would enable rigorous dose calculation in radiotherapy planning as well as more accurate segmentation and volume-calculation of organs \cite{3}. Another example is PET attenuation correction in PET/MRI. In this case, auto-completed MR scans can be utilized to calculate the attenuation coefficients rather than for diagnostic purposes.


Whereas medical image inpainting has relied so far on classical approaches (interpolation \cite{4,5} and diffusion techniques \cite{6}), the deep learning community has developed different frameworks to tackle this problem for natural images utilizing Generative Adversarial Networks (GANs) \cite{7}. One foundational and widely used framework is Context Encoder (CE) which consists of an auto-encoder structure regularized via an additional discriminator network \cite{8}. However, this framework only tackles the problem of inpainting a single square-shaped mask. It results in inpainted regions which do not always fit homogeneously into the surrounding context information. Another inpainting framework is the Globally and Locally Consistent Image Completion (GLCIC) framework \cite{9}. It enables inpainting of arbitrary shaped regions by expanding the CE framework to utilize two discriminator networks. However, further post-processing and long training times are required to ensure consistency of the inpainted region with the surrounding image. More recently, a GAN architecture based on gated convolutions was proposed for the inpainting of arbitrary shaped regions with strong competitive results \cite{10}. Despite this, the framework demands accurate localization of the pixel locations in which inpainting must be performed. This must be done either using a segmentation network during pre-processing or on-the-fly during training resulting in increased training complexity.

In our previous work, we introduced the ip-MedGAN framework for the inpainting of medical modalities, e.g. MRI and CT \cite{11}. This framework surpassed the CE and GLCIC frameworks both qualitatively and quantitatively. However, it suffered from two main drawbacks. First, it is restricted to only square-shaped inpainting regions. Second, it also demands exact localization of the masked region of interest.

In this work, we present a new framework for the inpainting of medical modalities, named ipA-MedGAN. The proposed framework bypasses the limitations of ip-MedGAN by enabling the inpainting of arbitrary shaped regions. Unlike gated convolutions, no localization of the arbitrary masked regions is required in advance. Experiments were conducted on brain MR scans as an example of the applicability of the proposed approach. A thorough qualitative and quantitative comparative analysis is performed between the proposed framework and other inpainting frameworks and image translation approaches for both arbitrary and square-shaped inpainting.

\begin{figure*}[!t]
	\centering
	
	\includegraphics[width=.85\textwidth]{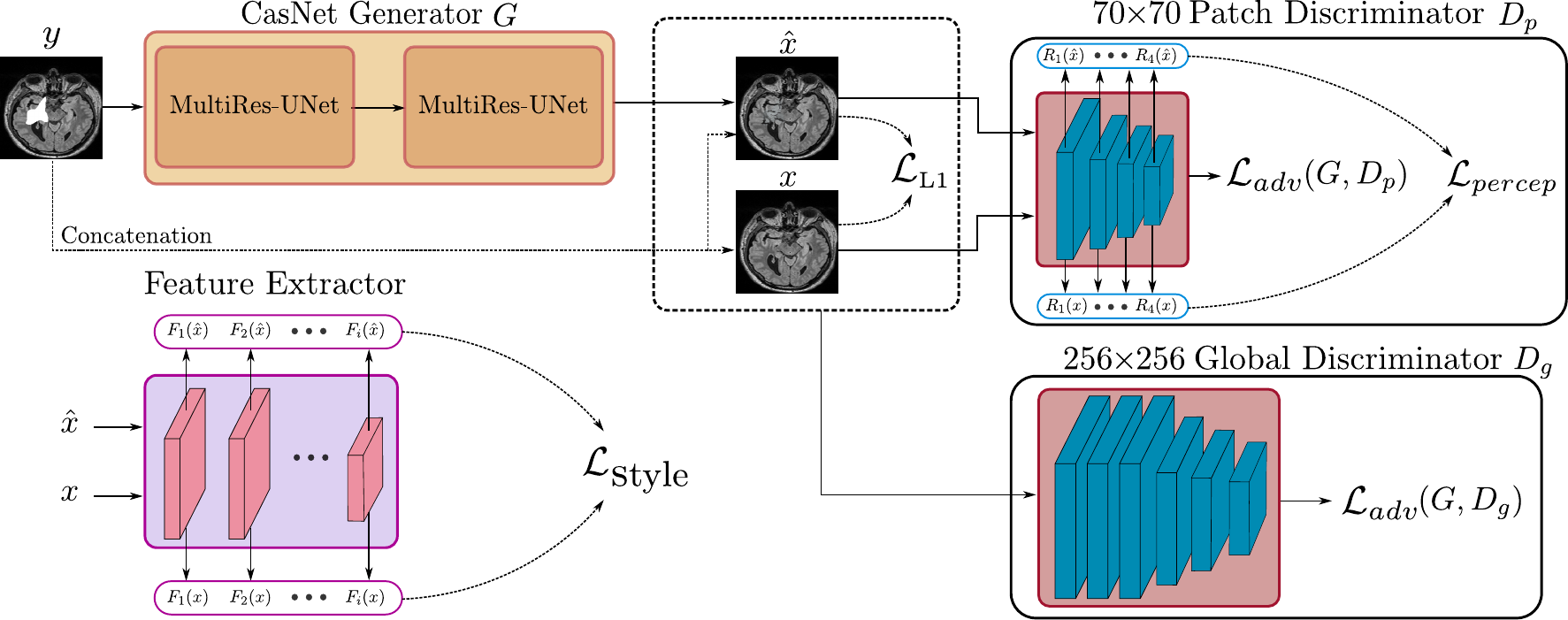}
	
	\caption{An overview of the proposed ipA-GAN framework for the inpainting of medical image modalities.}
	\label{fig1}
	\vspace{-2mm}
\end{figure*}

\section{Methods}

The proposed ipA-MedGAN is constructed using a new cascaded generator network, based on MultiRes-UNets \cite{12}, and two discriminator networks trained jointly as a conditional GAN (cGAN) \cite{13}. A pre-trained feature extractor network is utilized for the calculation of an additional non-adversarial loss. An overview of the proposed ipA-MedGAN framework is presented in Fig.~\ref{fig1}.\vspace{-2mm}

\subsection{Conditional GANs}
A cGAN is an extended version of GAN where the generator $G$ and the discriminator $D$  are conditioned on the input source image $y$  \cite{13}. In our case, $y$ is a 2D medical image of size $256 \times 256$ pixels with an arbitrary-shaped cropped region. $G$ is tasked with mapping the input to the desired output modality $G(y) = \hat{x}$, i.e. an inpainted image. This is guided by $D$, which acts as a binary classifier attempting to classify the ground-truth target image $x$ and the generator output $\hat{x}$ as real and fake, respectively. The two networks are trained in competition where $G$ attempts to fool $D$ into making a false decision. This is represented by the min-max optimization task over the adversarial loss function given as
\begin{equation}
\begin{split}
\min_{G} \max_{D} \mathcal{L}_{\small\textrm{adv}}(G,D) = & \; \mathbb{E}_{y} \left[\textrm{log} D(x,y) \right] + \\
& \; \mathbb{E}_{x} \left[\textrm{log} \left( 1 - D\left(\hat{x},y\right) \right) \right]
\end{split}
\end{equation}
\subsection{The Inpainting Set-up}

The previous ip-MedGAN framework relies on a local discriminator architecture which receives as input only the inpainted regions of interest instead of the full output $\hat{x}$ and target images $x$. This demands a prior localization of the masked regions, e.g. via a segmentation network, and restricts the inpainting process to a set of limited shapes. To counteract this problem, ipA-MedGAN utilizes a combination of two discriminator networks with different receptive fields. The first network, the global discriminator $D_g$, consists of 6 convolutional layers with a receptive field of $256 \times 256$. With such a wide receptive field, this network focuses on the global properties of the resultant images, thus ensuring the inpainted regions fit homogeneously into the surrounding context information. The second network, the patch discriminator $D_p$, is a shallower network of 4 convolutional layers and a $70 \times 70$ receptive fields. It can be viewed as dividing the input images into overlapping patches of size $70 \times 70$ pixels, classifying each patch and averaging out the classification scores for the final outcome. By focusing on smaller overlapping image regions, this network warrants a higher level of details within the inpainted regions. The detailed architectural design of both discriminator networks is provided in architectural analysis provided in \cite{13}. The loss function for this cGAN set-up is a combination of each respective set of adversarial loss functions:
\begin{equation}
\setlength\abovedisplayskip{2pt}
 \setlength\belowdisplayskip{0pt}
 \mathcal{L}_{\small\textrm{adv}} = \mathcal{L}_{\small\textrm{adv}}(G,D_g) + \mathcal{L}_{\small\textrm{adv}}(G,D_p)
\end{equation}

\subsection{Non-Adversarial Losses}

Additional non-adversarial losses are utilized to further regularize the generator network to achieve better results. The first of such losses is the perceptual loss which minimize the discrepancies between the resultant and the target images in the feature-space to produce globally consistent images \cite{14}. For this purpose, intermediate feature-maps are extracted from $D_p$ for both sets of images, $x$ and $\hat{x}$, and the mean absolute error (MAE) is calculated. Thus, the perceptual loss is expressed as:
 \begin{equation}
\mathcal{L}_{\small\textrm{percep}} = \sum_{i = 1}^{4} \lambda_{pi} \lVert{R\textsubscript{i}\left(\hat{x}\right) - R\textsubscript{i}\left(x\right)}\rVert_1
\end{equation}
where $R_i$ and $\lambda_{pi} > 0$ are the extracted feature-map and the weight of $i^{\textrm{th}}$ discriminator layer, respectively. A conventional pixel-reconstruction loss ($\mathcal{L}_{\small\textrm{L1}}$) is also utilized by minimizing the MAE between the raw inputs.

The second utilized non-adversarial loss is the style-reconstruction loss inspired from works on neural style-transfer \cite{15,155}. It aims to enhance textural details of the inpainted regions. This is achieved by using the feature-maps extracted from a pre-trained feature extractor network, in this case a VGG-19 network trained on the ImageNet classification task \cite{16}. The correlation between the feature-maps in the spatial extend, represented by the Gram matrices $Gr_n(x)$ and $Gr_n(\hat{x})$, is first calculated with final loss given as:\vspace{-2mm}
 \begin{equation}
 \vspace{-1mm}
\mathcal{L}_{\small\textrm{style}} = \sum_{n = 1}^{6} \lambda_{sn} \frac{1}{4 d_n^2} \lVert{Gr_n\left(\hat{x}\right) - Gr_n\left(x\right)}\rVert_F^{2}
\end{equation}
where $\lambda_{sn} > 0$ and $d_n$ are the weight and the spatial depth of the extracted feature-maps from the $n^{\textrm{th}}$ layer of the pre-trained feature extractor network, respectively \cite{155}.
\vspace{-2mm}
\subsection{MultiResUNet based Generator}
\begin{figure}[!t]
	\centering
	
	\includegraphics[width=0.95\columnwidth]{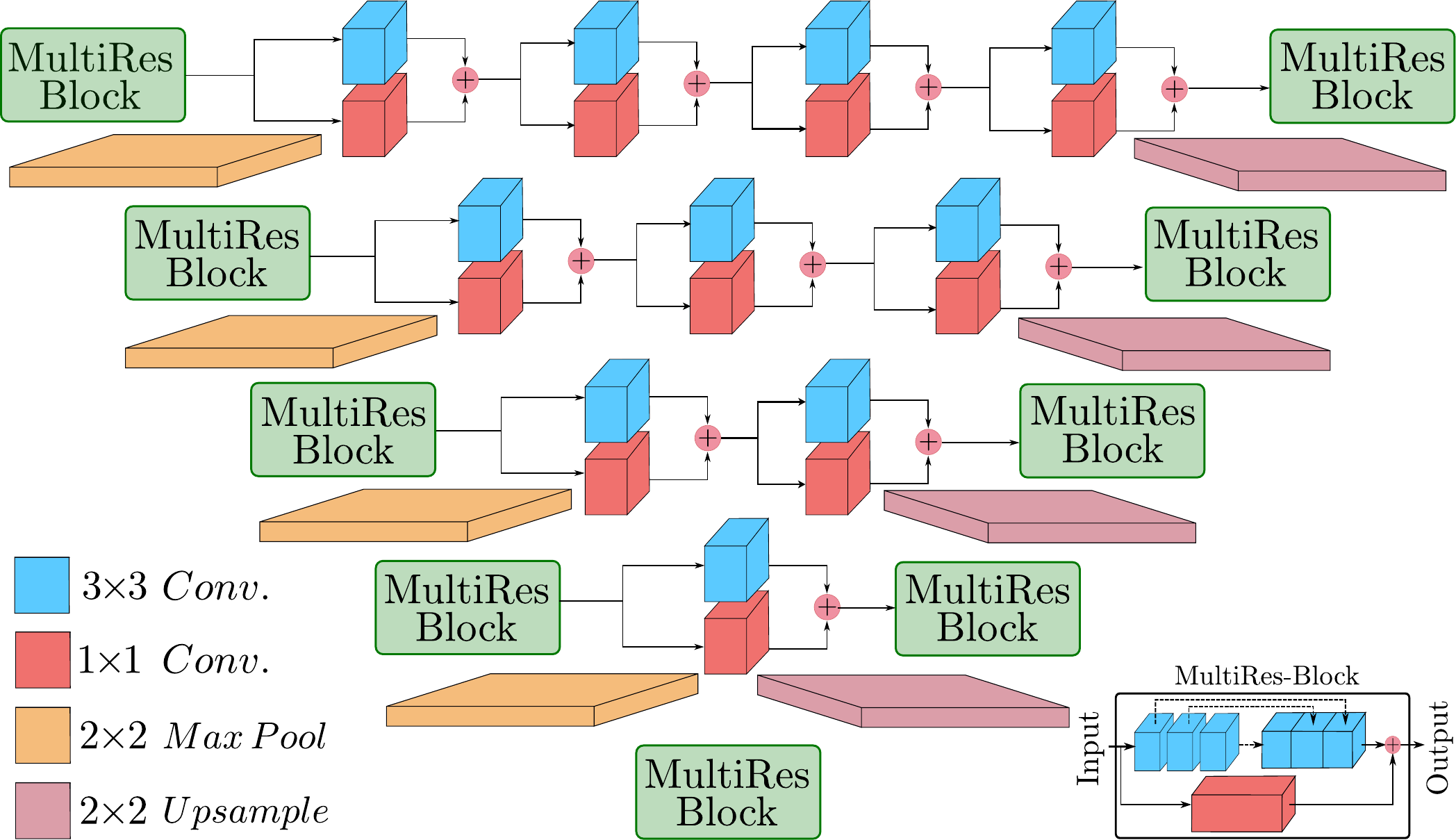}
	
	\caption{An overview of the MultiRes-Unet.}
	\label{fig2}
	\vspace{-2mm}
\end{figure}

A cascaded generator architecture which consists of two MultiRes-UNet is utilized \cite{12}. MultiRes-UNet is a state-of-the-art architectural design recently introduced as an improvement upon the classical U-Net architecture \cite{19}. It replaces every individual encoder-decoder level of the legacy U-net architecture with a MultiRes-Block. Each block consists of three cascaded $3 \times 3$ convolutional layers interconnected together to draw out various scales of spatial features. A  $1 \times 1$ convolution is added as a residual connection from the input of the MultiRes-Block to the output, in order to append the spatial information. To attenuate the differences in the feature-representations between encoder-decoder paired layers, shortcut connections, known as ResPaths, are added which consists of $3 \times 3$ and $1 \times 1$ convolutional filters. The architecture of the MultiRes-UNet is illustrated in Fig.~\ref{fig2}.

To summarize, the final loss function of ipA-MedGAN is the combination of adversarial and non-adversarial losses:
\begin{equation}
\mathcal{L} = \lambda_1 \mathcal{L}_{\small\textrm{adv}} +  \lambda_2 \mathcal{L}_{\small\textrm{style}} + \lambda_3 \mathcal{L}_{\small\textrm{percep}}
\end{equation}
where $\lambda_{1}$, $\lambda_{2}$ and $\lambda_{3}$ represent the contributions of the different loss functions. These hyperparameters were determined using the Bayesian optimization technique provided in \cite{20}.

\begin{table}[!h]
	\caption{Quantitative comparisons.\label{t1}}
	\centering
	\vspace{-2mm}
	\setlength\arrayrulewidth{0.05pt}
	\small
	\bgroup
	\def\arraystretch{1.15}
	\resizebox{\columnwidth}{!}{%
		\begin{tabular}{r|cccc}
			\hline\hline
			\multirow{2}{*}{Model} & \multicolumn{4}{c}{(a) Square shaped regions}\\ & SSIM & PSNR(dB) & MSE & UQI\\
			\hline
			CE & 0.8556 & 23.79 & 281.59 & 0.8006\\
			ip-MedGAN & 0.9518 & 29.68 & 81.14 & 0.8628\\
			ipA-MedGAN & \textbf{0.9606} & \textbf{30.62} & \textbf{65.81} & \textbf{0.9620}
			\\
			\hline \hline
			\multirow{2}{*}{Model} & \multicolumn{4}{c}{(b) Arbitrary shaped regions}\\ & SSIM & PSNR(dB) & MSE & UQI\\
			\hline
			pix2pix & 0.9667 & 32.34 & 43.77 & 0.8379\\
			MedGAN & 0.9653 & 31.92 & 47.42 & 0.8732\\
			Gated-Conv & 0.9650 & 31.46 & 50.26 & 0.8094
			\\
			ipA-MedGAN & \textbf{0.9818} & \textbf{35.12} & \textbf{24.90} & \textbf{0.9889}
			\\
			\hline
		\end{tabular}
	}
	\egroup
	\vspace{-5mm}
\end{table}

	\begin{figure*}
	\centering
	{\resizebox{\textwidth}{!}{\input{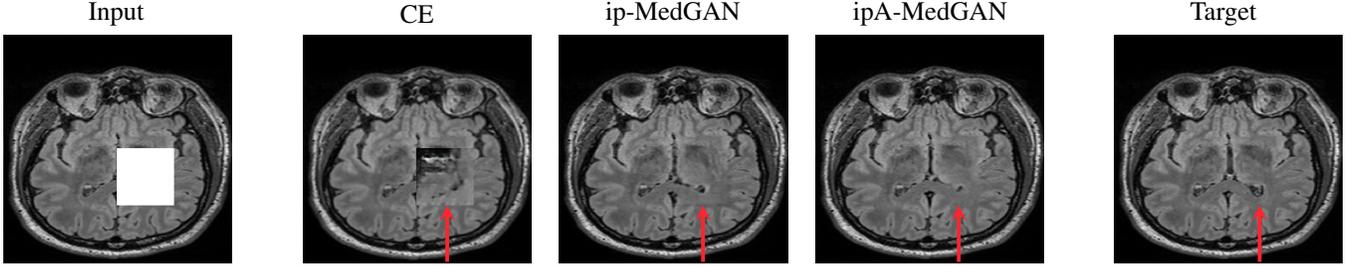}}}
	\caption{Qualitative comparison of inpainting for square-shaped regions with (\myarrowred) showing the advantages of ipA-MedGAN.}
	\label{fig3}
	\vspace{-5mm}
    \end{figure*}
	
	\begin{figure*}
	\centering
	\vspace{4mm}
	{\resizebox{\textwidth}{!}{\input{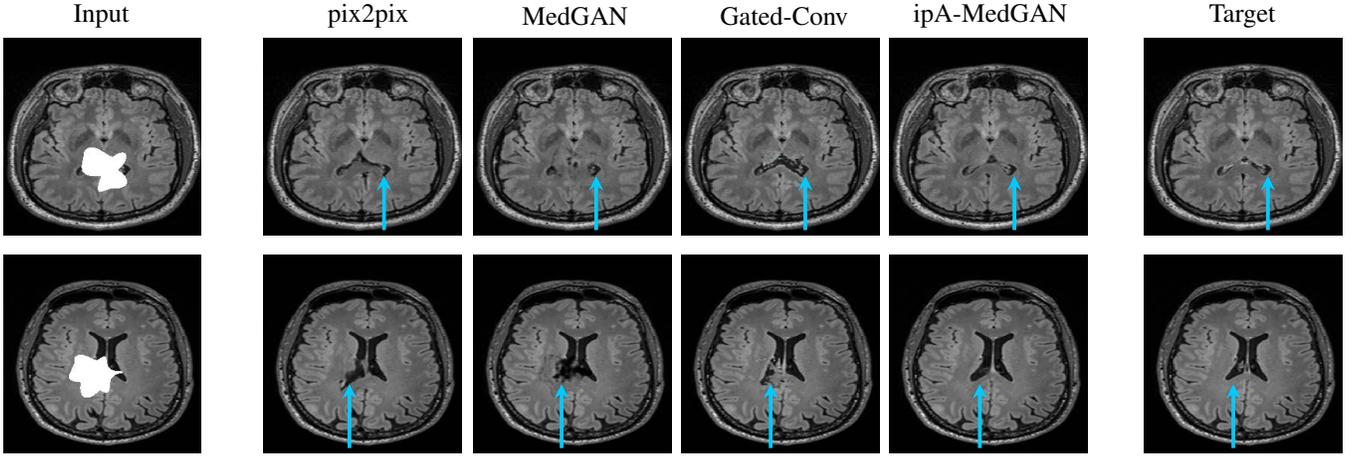}}}
	\caption{Qualitative comparison of inpainting for arbitrary-shaped regions with (\myarrowblue) showing the advantages of ipA-MedGAN.}
	\label{fig4}
	\vspace{-4.1mm}
\end{figure*}
\vspace{-2mm}
\section{Datasets and Experiments}
\vspace{-1mm}
The proposed ipA-MedGAN framework is evaluated on T2-weighted (FLAIR) MR dataset from the brain region. The MR data was acquired from 44 anonymized volunteers using a 3T scanner. The data was resampled to $1 \times 1 \times 1\textrm{mm}^{3}$ and rescaled to 2D slices of size $256 \times 256$ pixels. For training and testing, 3028 slices from 33 volunteers and 1072 slices from 11 volunteers were utilized, respectively. 

Two distinct set experiments were conducted. The first is the inpainting of traditional square-shaped masks. For this purpose, square-shaped regions of size $64 \times 64$ pixels (6.25 \% of the total image area) were randomly cropped from the pre-processed scans. The performance of ipA-MedGAN was compared against CE \cite{8} and ip-MedGAN \cite{11}, both specifically designed for square-shaped inpainting.

The second set of experiments focuses on the inpainting of arbitrary shaped regions. To achieve this, a set of 100 random masks, whose size range from 1.36\% to 5.46\% of the total image area, were utilized to create the input MR scans $y$. For validation, a different set of 50 masks, whose size range from 1.66\% to 4.56\%, was utilized to test the generalization capability of the proposed framework. To evaluate the performance of ipA-MedGAN, qualitative and quantitative comparisons were conducted against the gated convolutions inpainting framework (Gated-Conv) \cite{10} and other image-translation approaches, i.e. pix2pix \cite{13} and MedGAN \cite{17}. 

To ensure a faithful representation of the approaches utilized in the comparative analysis, verified open-source implementations along with the hyperparameter settings from the original publications were utilized \cite{22,23}. All models were trained for 100 epochs. For the quantitative comparisons, several metrics were utilized: the Universal Quality Index (UQI) \cite{24}, Structural Similarity Index Measure (SSIM) \cite{25}, Peak Signal to Noise Ratio (PSNR) and the Mean Squared Error (MSE). Unlike the comparisons done in ip-MedGAN, all the quantitative metrics in this work are calculated on the entire output images (256 $\times$ 256 pixels) and not solely on the inpainted regions. This is reflected by the higher metric scores compared to the scores reported in ip-MedGAN.

\section{Results and Discussion}
\vspace{-2mm}
For square-shaped inpainting, the results are presented in Fig.~\ref{fig3} and Table~I~(a), respectively. 
CE resulted in the worst inpainting results from a qualitative perspective with the resultant inpainting regions not fitting homogeneously into the surrounding context information. The ip-MedGAN framework enhanced the inpainting quality both qualitatively and quantitatively. However, as illustrated by (\myarrowred) in Fig.~\ref{fig3}, minor visual artifacts can still be depicted at the boundary of the inpainted regions. Additionally, the exact location of the region of interest must be supplied as input to ip-MedGAN during training. The proposed ipA-MedGAN overcomes both of the previous drawbacks by eliminating the boundary artifacts with no required prior localization of the masked regions.

In Fig.~\ref{fig4} and Table~I~(b), the qualitative and quantitative results for arbitrary inpainting of MR are displayed, respectively. The gated convolution framework produced inpainting regions with altered textural and anatomical characteristics when compared to the target images. This was also indicated quantitatively by the inferior scores in Table~I~(b). Generative image-to-image translation approaches resulted in better qualitative results that fit more homogeneously with the surrounding regions. However, the inpainted images produced by pix2pix depicted blurry and unrealistic textures. On the other hand, results by MedGAN contained distinct visual tilting artifacts. The proposed ipA-MedGAN framework resulted in more coherent and artifact-free inpainted regions which closely matche the desired target images. This was also reflected quantitatively with the scores in Table~I~(b). More notably, the significantly improved UQI and MSE scores. Compared to other traditional inpainting frameworks such as gated convolutions and GLCIC, this comes with the additional benefit of not requiring prior localization of the regions of interest during training.

This work is not without limitations. In the future, we must extend the current work to suit 3D medical scans in order to incorporate the correlation information in the spatial extent. Furthermore, we plan to conduct investigations of the validity of the resultant images in clinical post-processing tasks and real-world MRI processing applications.

\vspace{-3mm}
\section{Conclusion}
\vspace{-2mm}
In this work, a new framework for the inpainting of medical modalities with local distortions is introduced. The ipA-MedGAN framework overcomes the limitations of previous approaches by enabling the inpainting of arbitrary shaped regions with no a priori knowledge on the pixel locations of the regions of interest. This is achieved by using a combination of a cascaded MultiRes-UNets generator, two discriminator networks with different receptive fields and a pre-trained feature extractor network. A comparative analysis with other inpainting approaches have illustrated the superior performance of ipA-MedGAN for the inpainting of brain MR images. 

%

%


\newpage
\bibliographystyle{IEEEbib}
\balance
{\normalsize
	\bibliography{refs2}}

\begin{thebibliography}{10}

\bibitem{1}
K.~{Armanious}, S.~{Gatidis}, K.~{Nikolaou}, B.~{Yang}, and T.~{Kustner},
\newblock ``Retrospective correction of rigid and non-rigid mr motion artifacts
  using gans,''
\newblock in {\em IEEE 16th International Symposium on Biomedical Imaging
  (ISBI)}, April 2019, pp. 1550--1554.

\bibitem{2}
K.M. Koch, B.A. Hargreaves, K.~Butts Pauly, W.~Chen, G.E. Gold, and K.F. King,
\newblock ``Magnetic resonance imaging near metal implants,''
\newblock {\em Journal of Magnetic Resonance Imaging}, vol. 32, no. 4, pp.
  773--787, 2010.

\bibitem{3}
M.~Arnold, A.~Ghosh, S.~Ameling, and G.~Lacey,
\newblock ``Automatic segmentation and inpainting of specular highlights for
  endoscopic imaging,''
\newblock {\em Journal on Image and Video Processing}, 2010.

\bibitem{4}
M.~Alsalamah and S.~Amin,
\newblock ``Medical image inpainting with {RBF} interpolation technique,''
\newblock {\em International Journal of Advanced Computer Science and
  Applications}, vol. 7, 2016.

\bibitem{5}
Z.~Feng, S.~Chi, J.~Yin, D.~Zhao, and X.~Liu,
\newblock ``A variational approach to medical image inpainting based on
  mumford-shah model,''
\newblock in {\em International Conference on Service Systems and Service
  Management}, 2007.

\bibitem{6}
P.~Vlašánek,
\newblock ``Fuzzy image inpainting aimed to medical imagesl,''
\newblock in {\em International Conference in Central Europe on Computer
  Graphics, Visualization and Computer Vision}, 2018.

\bibitem{7}
I.~J. Goodfellow, J.~Pouget-Abadie, M.~Mirza, B.~Xu, D.~Warde-Farley, S.~Ozair,
  A.~C. Courville, and Y.~Bengio,
\newblock ``Generative adversarial nets,''
\newblock in {\em Conference on Neural Information Processing Systems}, 2014,
  pp. 2672--2680.

\bibitem{8}
D.~Pathak, P.~Kr{\"a}henb{\"u}hl, J.~Donahue, T.~Darrell, and A.~A. Efros,
\newblock ``Context encoders: Feature learning by inpainting,''
\newblock {\em IEEE Conference on Computer Vision and Pattern Recognition
  (CVPR)}, pp. 2536--2544, 2016.

\bibitem{9}
S.~Iizuka, E.~Simo-Serra, and H.~Ishikawa,
\newblock ``{Globally and Locally Consistent Image Completion},''
\newblock {\em ACM Transactions on Graphics (Proc. of SIGGRAPH 2017)}, vol. 36,
  2017.

\bibitem{10}
J.~Yu, Z.~Lin, J.~Yang, X.~Shen, X.~Lu, and T.~S Huang,
\newblock ``Free-form image inpainting with gated convolution,''
  \url{https://arxiv.org/abs/1806.03589}, 2018,
\newblock arXiv preprint.

\bibitem{11}
K.~{Armanious}, Y.~{Mecky}, S.~{Gatidis}, and B.~{Yang},
\newblock ``Adversarial inpainting of medical image modalities,''
\newblock in {\em IEEE International Conference on Acoustics, Speech and Signal
  Processing (ICASSP)}, May 2019, pp. 3267--3271.

\bibitem{12}
N.~Ibtehaz and M.~S. Rahman,
\newblock ``{MultiResUNet : Rethinking the U-Net architecture for multimodal
  biomedical image segmentation},''
\newblock {\em Neural Networks}, vol. 121, pp. 74 -- 87, 2020.

\bibitem{13}
P.~Isola, J.~Zhu, T.~Zhou, and A.~A. Efros,
\newblock ``Image-to-image translation with conditional adversarial networks,''
\newblock in {\em Conference on Computer Vision and Pattern Recognition}, 2016,
  pp. 5967--5976.

\bibitem{14}
C.~Wang, C.~Xu, C.~Wang, and D.~Tao,
\newblock ``Perceptual adversarial networks for image-to-image
  transformation,''
\newblock {\em IEEE Transactions on Image Processing}, vol. 27, 2018.

\bibitem{15}
L.~A. Gatys, A.~S. Ecker, and M.~Bethge,
\newblock ``Image style transfer using convolutional neural networks,''
\newblock in {\em IEEE Conference on Computer Vision and Pattern Recognition},
  2016, pp. 2414--2423.

\bibitem{155}
J.~Johnson, A.~Alahi, and F.~Li,
\newblock ``Perceptual losses for real-time style transfer and
  super-resolution,''
\newblock 2016, pp. 694--711.

\bibitem{16}
K.~Simonyan and A.~Zisserman,
\newblock ``Very deep convolutional networks for large-scale image
  recognition,'' \url{http://arxiv.org/abs/1409.1556}, 2014,
\newblock arXiv preprint.

\bibitem{17}
K.~Armanious, C.~Yang, M.~Fischer, T.~K\"ustner, K.~Nikolaou, S.~Gatidis, and
  B.~Yang,
\newblock ``{MedGAN}: Medical image translation using {GANs},''
  \url{http://arxiv.org/abs/1806.06397v1}, 2018,
\newblock arXiv preprint.

\bibitem{19}
O.~Ronneberger, P.~Fischer, and T.~Brox,
\newblock ``U-net: Convolutional networks for biomedical image segmentation,''
\newblock in {\em Medical Image Computing and Computer-Assisted Intervention
  (MICCAI)}, 10 2015, vol. 9351, pp. 234--241.

\bibitem{20}
FMFN,
\newblock ``{Bayesian optimizazion} implementation,''
  \url{https://github.com/fmfn/BayesianOptimization}.

\bibitem{22}
Narihiro Tada,
\newblock ``{CE} implementation,''
  \url{https://github.com/jazzsaxmafia/Inpainting}.

\bibitem{23}
Jiahui Yu,
\newblock ``{Gated convolutions} implementation,''
  \url{https://github.com/JiahuiYu/generative_inpainting}.

\bibitem{24}
Z.~Wang and A.~C. Bovik,
\newblock ``A universal image quality index,''
\newblock in {\em IEEE Signal Processing Letters}, March 2002, vol.~9, pp.
  81--84.

\bibitem{25}
Z.~Wang, A.~C. Bovik, H.~R. Sheikh, and E.~P. Simoncelli,
\newblock ``Image quality assessment: from error visibility to structural
  similarity,''
\newblock in {\em IEEE Transactions on Image Processing}, 2004, vol.~13, pp.
  600--612.

\end{thebibliography}
\end{document}